\documentclass{svjour3}                     
\smartqed  
\usepackage{graphicx}
\journalname{Few-Body Systems (FB20)}
\begin{document}

\title{
Solving Bethe-Salpeter equation for scattering states
\thanks{Presented at the 20th International IUPAP Conference on Few-Body Problems in Physics, 20 - 25 August, 2012, Fukuoka, Japan}
}


\author{V.A.~Karmanov     \and J.~Carbonell
}


\institute{V.A.~Karmanov \at
              Lebedev Physical Institute, Leninsky prospect 53, 119991 Moscow, Russia\\
              \email{karmanov@sci.lebedev.ru}
 \and
J.~Carbonell \at CEA-Saclay, IRFU/SPhN, 91191 Gif-sur-Yvette, France}

\date{Received: date / Accepted: date}

\maketitle

\begin{abstract}
We present the Minkowski space solutions of the inhomogeneous  Bethe-Salpeter equation for spinless particles with a ladder kernel.
The off-mass shell scattering amplitude is first obtained.
\end{abstract}

\section{Introduction}
\label{intro}
The inhomogeneous Bethe-Salpeter (BS) equation in Minkowski space \cite{bs}
\begin{small}
\begin{equation}\label{BSE}
F(p,p_s; P)=V(p,p_s; P)- i\int\frac{d^4p'}{(2\pi)^4}
\frac{V(p,p';P) F(p',p_s,P)}
{\left[\left(\frac{P}{2}+p'\right)^2-m^2+i\epsilon\right]
\left[\left(\frac{P}{2}-p'\right)^2-m^2+i\epsilon\right]}
\end{equation}
\end{small}
provides a covariant four-dimensional  description of two-body  scattering states. In the center of mass frame, and for a given incident  momentum $p_s$, the partial wave off-mass shell amplitude $F_l$ depends on two scalar variables $p_0$ and $|\vec{p}|$. It will be hereafter denoted by
$F_l(p_0,p;p_s)$   setting $p=|\vec{p}|$, $p_s=|\vec{p_s}|$.

The on-shell amplitude $F^{on}_l=F_l(p_0=0,p=p_s;p_s)$
determines the phase shift according to:
\begin{equation}\label{delta}
\delta_l=\frac{1}{2i}\log\Bigl(1+\frac{2ip_s }{\varepsilon_{p_s}} F^{on}_l\Bigr)
\end{equation}
with $\varepsilon_{p_s}=\sqrt{m^2+p_s^2}$.
The knowledge of this function in the entire domain of its arguments -- i.e. the off-shell amplitude -- is mandatory for some interesting physical applications, like for instance  computing  the transition e.m. form factor $\gamma^*d\to np$ or  solving the BS-Faddeev equations.
This quantity  has not  been obtained until now.

The numerical solution of the BS equation in Minkowski space is complicated by the existence of singularities in the amplitude as well as in the integrand of (\ref{BSE}). These singularities are integrable in the mathematical sense, due to $i\epsilon$ in the denominators of propagators,
but  their  integration is a quite delicate task and requires the use  of appropriate analytical as well as  numerical methods.

To avoid these singularities, the BS equation was first solved in Euclidean space. These solutions provided   on-shell quantities like  binding energies and  phase shifts \cite{tjon}. However we have shown \cite{ck-trento} that the Euclidean BS amplitude cannot be used to calculate electromagnetic form factors, since the corresponding integral does not allow the Wick rotation.
One therefore needs the BS amplitude in Minkowski space.

This amplitude has been computed for a separable kernel  (see \cite{burov} and references therein). For a field-theory  kernel -- the ladder and the cross ladder --   it  has been first obtained in our preceding works \cite{bs1,bs2} for the bound state problem. To this aim, we developed a method based on the Nakanishi integral representation of the BS amplitude.
A similar method for the scattering states has been proposed in \cite{fsv-2012} although  the numerical solutions are not yet available.

We present in this contribution a new method providing a direct solution of the original BS equation. It is based  on a scrutinized treatment of the singularities and allows us to compute the corresponding off-shell scattering amplitude in Minkowski space. We will give the low energy parameters in the case of spinless particles and ladder kernel.

\section{Method}
\label{sec1}
There are four sources of singularities in the  r.h.-side of the BS equation (\ref{BSE}) which are detailed below.

{\it (i)} The constituent propagators have two poles, each of them represented  as:
$$
\frac{1}{p'_0-a-i\epsilon}=PV\frac{1}{p'_0-a}+i\pi \delta(p'_0-a)
$$
where $PV$ means the principal value. In the product of four pole terms, the only non vanishing contributions come from the product of four PVs without delta-functions, from the terms with three PVs and one delta-function and from the term with two PVs and two delta's. After partial wave decomposition the 4D integral BS equation is reduced to a 2D one. Integrating over $p'_0$, we obtain  in addition to the 2D  part, a 1D integral over $p'$ and a non integrated term. The singularities due to the PVs are eliminated by subtractions.

{\it (ii)} The  propagator of the exchanged particle has the pole singularities which, after  partial wave decomposition, turn into logarithmic ones. Their positions are found analytically and the numerical integration over $p'_0$ variable is split into intervals between two consecutive singularities, namely:
$$\int_0^{\infty}[ \ldots ] \;  dp'_0= \int_0^{sing_1} [ \ldots ] \; dp'_0 +
\int_{sing_1}^{sing_2}[ \ldots ] \; dp'_0 + \int_{sing_2}^{sing_3}[ \ldots ] \;  dp'_0 +\dots
$$
Each of these integrals is made regular with an appropriate change of variable. We proceed in a similar way for the $p'$ integration.

{\it (iii)}  The inhomogeneous (Born) term is given by the ladder kernel and is also singular in both variables. The positions of these singularities are  analytically known.

{\it (iv)}  The amplitude $F$ itself has many singularities, among which the Born term contains the strongest ones. This makes difficult its representation on a basis of regular functions as well as its numerical integration.
To circumvent this difficulties we made the replacement
$F=V f$, where $f$ is a smooth function. After that, the singularities of the inhomogeneous term are canceled.
We obtain in this way a non-singular equation for  $f$ which we solved by standard methods.
\begin{figure}[h]
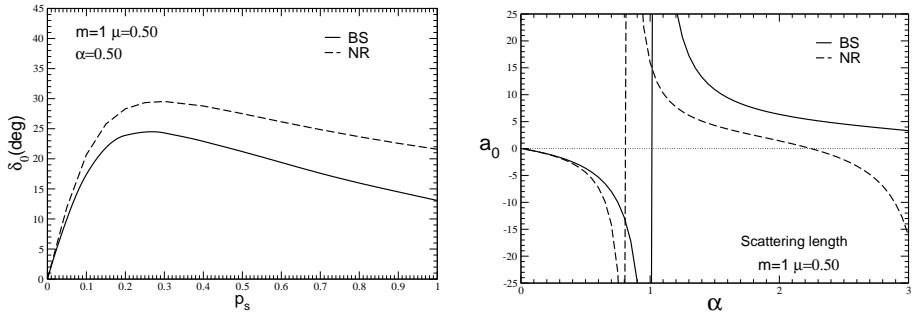

\centering
\includegraphics[width=5.7cm]{Phaseshifts_alpha_0.500_0.50_Black.eps}
\hspace{0.3cm}
\includegraphics[width=5.7cm]{a2_alpha_mu_0.50_Black.eps}
\caption{On left: phase shift calculated  via BS equation (solid curve) are compared to the non-relativistic results  (dashed curve) for $\mu=0.5$.
On right: the same comparison for the scattering length $a_0$ vs. the coupling constant  $\alpha$.}
\label{fig1}
\end{figure}

\section{Numerical results}
\label{sec2}

We first applied this method to solve the bound state problem by dropping the inhomogeneous term in (\ref{BSE}). The binding energies coincide, within four-digit accuracy, with the ones calculated in our previous work \cite{bs1} and with the Euclidean space results.

The S-wave off-shell scattering amplitude $F_0$ was calculated and the phase shifts extracted by means of eq. (\ref{delta}). They include an imaginary part, which has been also found, above the first inelastic threshold  $p^*_s(\mu)=\sqrt{m\mu+\mu^2/4}$. By performing a Wick rotation in
(\ref{BSE}) we derive an Euclidean space equation similar to one obtained in \cite{tjon}. The phase shifts found by these two methods coincide with each other within 3-4 digits. Furthermore, the imaginary part of the phase shifts vanishes with high accuracy below threshold. The unitarity condition is not automatically fulfilled in our approach, but appears as a consequence of handling the correct solution. It thus provides a stringent test of the numerical method. Our results reproduce the phase shifts given in  \cite{tjon} within the accuracy allowed by extracting numerical values from published figures.

Figure \ref{fig1}, left panel, shows the phase shifts calculated via BS equation (solid curve) and via the Schr\"odinger one with the Yukawa potential (dashed curve) for the constituent mass $m=1$,
exchange mass $\mu=0.5$ and coupling constant $\alpha=g^2/(16\pi m^2)=0.5$.
Right panel shows the same comparison for the scattering length $a_0$ as a function of  the coupling constant $\alpha$. In the vicinity of  $\alpha\approx 0.8$ (for Schr\"odinger) and $\alpha\approx 1$ (for BS) the coupling constant crosses the critical value corresponding to the appearance of a bound state. At this point the scattering length becomes infinite and then changes the sign. One can see that the differences between relativistic and non-relativistic results are rather large even for small and zero incident momenta. These differences  increase with the value of $\alpha$.

\begin{figure}
\centering
\includegraphics[width=5.7cm]{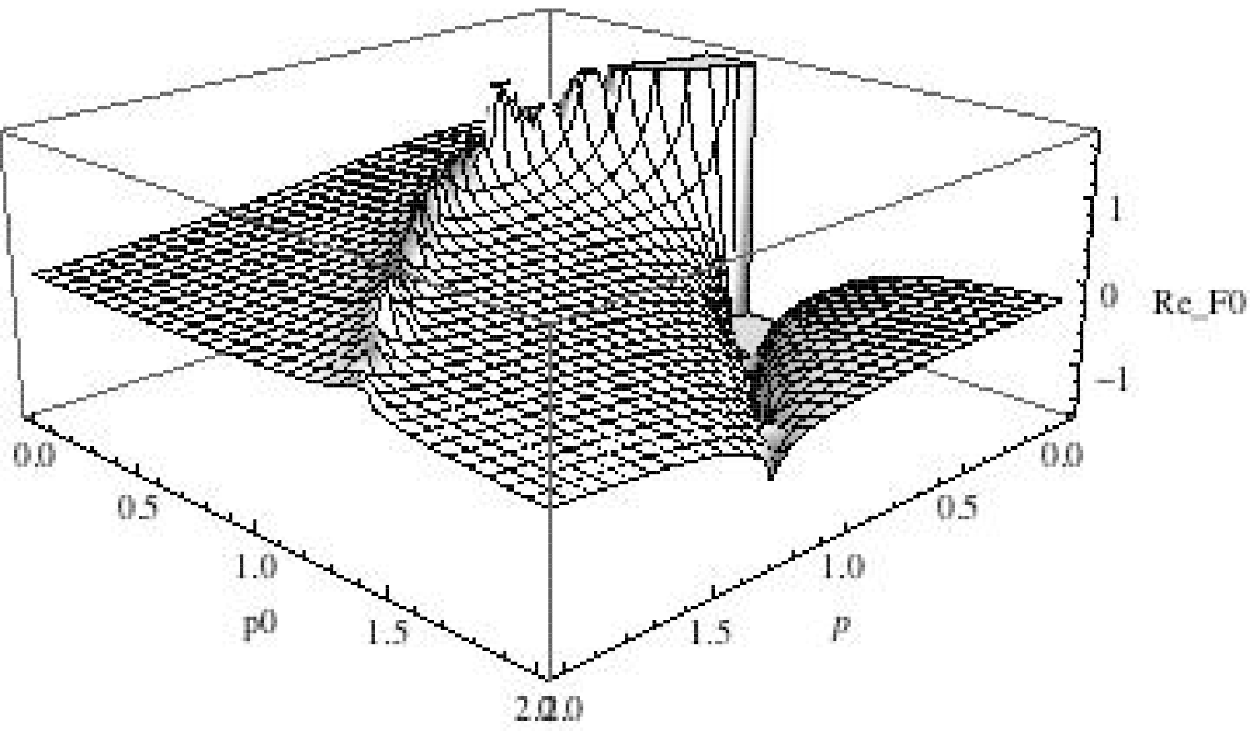}
\hspace{0.3cm}
\includegraphics[width=5.7cm]{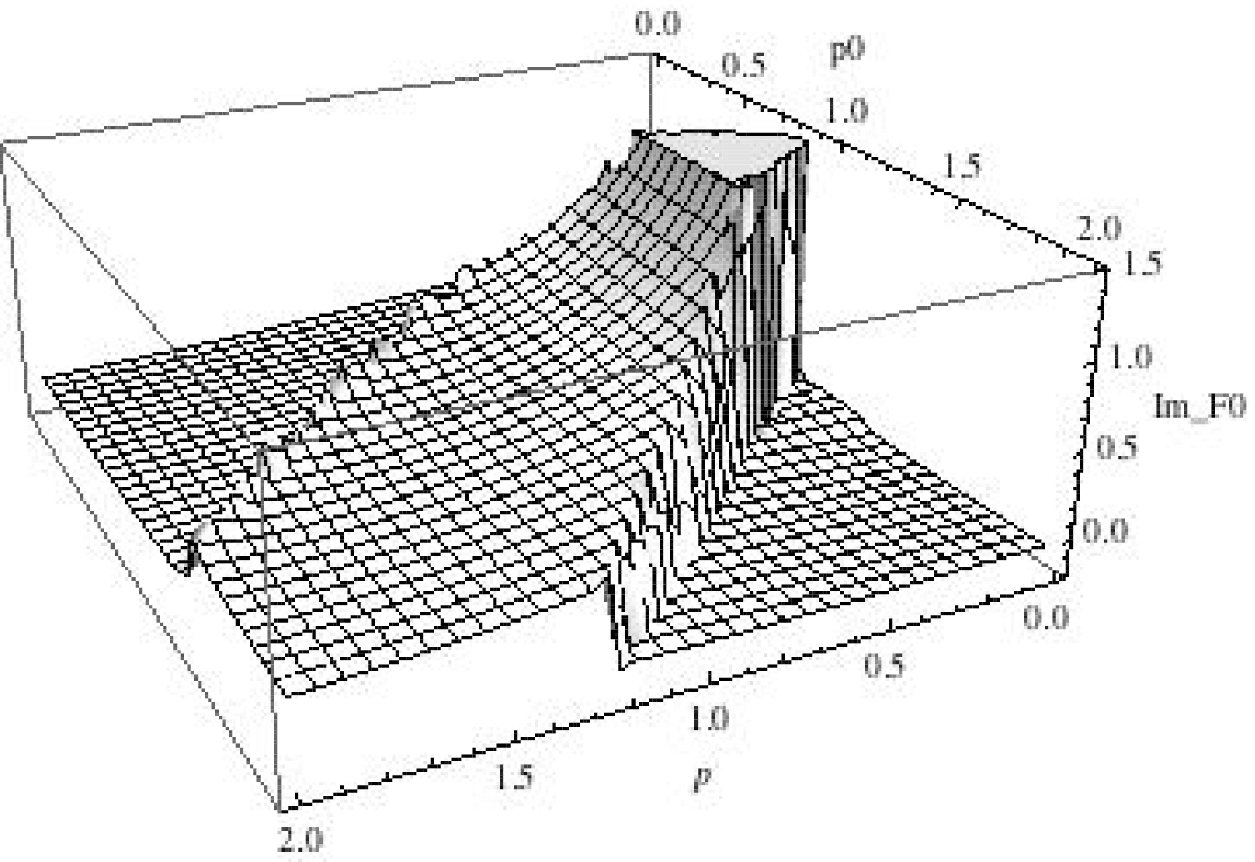}
\caption{On left: real part of the off-shell amplitude $F(p_0,p;p_s)$ for $p_s=0.5$, $\mu=0.5$.
On right: imaginary part of $F(p_0,p;p_s)$.}
\label{fig2}       
\end{figure}

We have displayed in Fig. \ref{fig2}  the real (left panel) and imaginary (right panel) parts of the off-shell scattering amplitude $F_0(p_0,p;p_s)$
vs. $p_0$ and $p$ calculated for $p_s=\mu=0.5$. Its real part shows a non trivial structure with  a ridge and a gap resulting from the singularities of the inhomogeneous term. Its on-shell value \mbox{$F_0^{on}=F_0(p_0=0,p=p_s;p_s)$}, determining the phase shift calculated previously, corresponds to a single point on theses surfaces. Our calculation, shown in Fig. \ref{fig2}, provides  the full amplitude $F_0(p_0,p;p_s)$ in  a two-dimensional domain.
Computing this quantity is the main result of this work.

\section{Conclusion}

We have presented the solutions of the BS equation for the scattering states in Minkowski space. Results  were limited to the spinless case, S-wave, and  ladder kernel. The full off-mass-shell amplitude is found for the first time. Its on-mass shell values provide the phase shifts. They are in agreement with those obtained in Euclidean space by other methods.
Relativistic phase shifts  considerably differ, even at low energy, from the non-relativistic ones calculated by the Schr\"odinger equation with a Yukawa potential. Our results include the inelasticities which appear above the meson creation threshold. The knowledge of the off-mass-shell amplitude in Minkowski space is necessary to calculate the transition form factor and in the three-body BS-Faddeev equations.


\end{document}